\documentclass[twocolumn,showpacs,prl,amsmath,amssymb]{revtex4}
\usepackage{graphicx,bm}
\usepackage{epsfig}

\begin{document}
\title{Spin projection in the shell model Monte Carlo method and the spin distribution of nuclear level densities}
\author{Y. Alhassid$^1$, S. Liu$^1$ and H. Nakada$^2$}
\affiliation{$^1$Center for Theoretical Physics, Sloane Physics Laboratory, Yale University,  New Haven, Connecticut 06520, U.S.A. \\
$^2$Department of Physics, Chiba University,
Inage, Chiba 263-8522, Japan}

\begin{abstract}
 We introduce spin projection methods in the shell model Monte Carlo approach and apply them to calculate the spin distribution of level densities for iron-region nuclei using the complete $(pf+g_{9/2})$-shell. We compare the calculated distributions with the spin-cutoff model and extract an energy-dependent moment of inertia. For even-even nuclei and at low excitation energies, we observe a significant suppression of the moment of inertia and odd-even staggering in the spin dependence of level densities.
\end{abstract}
\maketitle

The spin distribution of level densities are important for the calculation of statistical nuclear reaction rates such as those in thermal stellar reactions \cite{rauscher00}. Knowledge of the spin distribution is also required for the determination of total level densities from measured neutron or proton resonances \cite{di73,il92}, since the latter are subjected to spin selection rules.

The microscopic calculation of the spin distribution of level densities in the presence of correlations is a difficult problem. It is often assumed that the spin distribution follows the spin-cutoff model, obtained in the random coupling model of uncorrelated spins of the individual nucleons or excitons~\cite{er60}. The spin-cutoff distribution is determined by a single parameter, an effective moment of inertia. The latter is often set to its rigid-body value and occasionally determined empirically.

The interacting shell model takes into account both shell effects and correlations and thus provides a suitable framework for the calculation of level densities. However, in mid-mass and heavy nuclei, the required model space is many orders of magnitude larger than spaces in which conventional diagonalization methods can be applied. This problem was overcome by using the shell model Monte Carlo (SMMC) approach~\cite{la93,al94} to calculate level densities~\cite{na97,al99}.
SMMC level densities in the iron region were found to be in good agreement with experimental data without any adjustable parameters \cite{na97,al99}.

In the SMMC approach, thermal averages are taken over all possible
states of a given nucleus, and thus the computed level densities are
are those summed over all possible spin values. Here we introduce spin projection
methods within the SMMC approach that enable us to calculate thermal
observables at constant spin.
We first discuss projection on a given spin component $J_z$,
and then use it to calculate spin-projected expectation values of {\em scalar} observables.

We apply the method to the spin distribution of level densities in
the iron region, and compare the results with the spin-cutoff model.
We also extract from the spin distributions an energy-dependent moment
of inertia.
Signatures of the pairing phase transition are observed
in the energy dependence of the moment of inertia.

We first introduce a projection on the spin component $J_z=M$ along a fixed $z$-axis.  The projected partition function for a fixed value of $M$ and at inverse temperature $\beta$ is defined by  $Z_M(\beta) \equiv  {\rm Tr}_M e^{-\beta H}$ where $H$ is the Hamiltonian of the system and
\begin{equation}
\label{M-projection}
{\rm Tr}_M \hat X \equiv \sum\limits_{\alpha, J \ge
|M|} \langle \alpha J M | \hat X | \alpha J M \rangle 
\end{equation}
for an operator $\hat X$.
Here we assumed $H$ to be rotationally invariant, so its eigenstates $|\alpha J M\rangle$ are characterized by good total angular momentum $J$ and its magnetic quantum number $M$ with $M$-independent energies $E_{\alpha J}$.  The label $\alpha$ distinguishes between states with the same spin $J$.  The $M$-projected partition is then given by $Z_M(\beta) = \sum_{\alpha, J \ge |M|} e^{-\beta E_{\alpha J}}$.
In the following we assume all traces to be canonical, i.e., at fixed number $Z,N$ of protons and neutrons, unless otherwise stated.

 The Monte Carlo method is based on the Hubbard-Stratonovich (HS) transformation $e^{-\beta  H} = \int D[\sigma] G_\sigma U_\sigma$, where $G_\sigma$ is a
Gaussian  weight and $U_\sigma$ is the imaginary-time propagator of non-interacting nucleons moving in auxiliary fields $\sigma$. Using the HS representation,  the probability to find a state with a given spin projection $M$ at temperature $\beta^{-1}$ is
\begin{equation}\label{M-ratio}
{Z_M(\beta) \over Z(\beta)} = { \left\langle {{\rm Tr}_M U_\sigma \over {\rm  Tr} U_\sigma} \Phi_\sigma \right\rangle_W  \over \langle \Phi_\sigma \rangle_W}
\;,
\end{equation}
where we have introduced the notation
$\langle X_\sigma \rangle_W \equiv {\int D[\sigma] W(\sigma) X_\sigma /
\int  D[\sigma] W(\sigma)}$,
and $W(\sigma)\equiv G_\sigma |{\rm Tr} U_\sigma |$ is a positive-definite function used for the Monte Carlo sampling.
$\Phi_\sigma = {\rm Tr} U_\sigma  /|{\rm Tr} U_\sigma |$ in (\ref{M-ratio}) is the Monte Carlo sign.

In general $U_\sigma$ is not rotationally invariant, and the
$M$-projected partition ${\rm Tr}_M U_\sigma$
can be calculated by $J_z$ projection.  To this end, we use the identity
\begin{equation}
  \label{fourier}
  {\rm Tr}\,(e^{i\varphi_k \hat J_z} U_\sigma) = \sum\limits_{M=-J_s}^{J_s} e^{i\varphi_k M} {\rm Tr}_M\,U_\sigma\;,
\end{equation}
where $J_s$ is the maximal many-particle spin in the model space and  $\varphi_k$ assumes a discrete set of values. Using the $2J_s+1$ quadrature points $\varphi_k \equiv \pi {k \over J_s+1/2}$ ($k=-J_s, \ldots J_s$),  the set of discrete functions $\chi_M (\varphi_k) \equiv (2J_s+1)^{-1/2} e^{i\varphi_k M}$ is orthonormal,
$\sum_{k = -J_s}^{J_s} \chi_M(\varphi_k)\chi_{M'}^\ast(\varphi_k)  = \delta_{M M'}$. This orthogonality relation can
 be used to invert (\ref{fourier})
\begin{equation}
  \label{M-project}
  {\rm Tr}_M \,U_\sigma = {1 \over 2J_s + 1} \sum\limits_{k =-J_s}^{J_s}
  e^{-i \varphi_k M} {\rm Tr}\,\left( e^{i\varphi_k \hat J_z} U_\sigma \right) \;.
\end{equation}

The trace on the r.h.s. of (\ref{M-project}) is a canonical trace at a
fixed particle number ${\cal A}$ (in practice we need to project on both $N$ and $Z$),
and is calculated from the grand-canonical traces by a particle-number projection
\begin{equation}\label{canonical-trace}
{\rm Tr}\,\left( e^{i\varphi_k \hat J_z} U_\sigma \right)= {1\over N_s} \sum\limits_{n=1}^{N_s} e^{-i \chi_n{\cal A}} \det ({\bf I} + {\bf U}^{(n,k)}_\sigma) \;.
\end{equation}
Here $\chi_n=\pi n/2N_s$ are quadrature points, and
${\bf U}^{(n,k)}_\sigma \equiv e^{i \chi_n} e^{i\varphi_k \hat j_z} {\bf U}_\sigma$
is the $N_s\times N_s$ matrix representing the many-particle propagator $e^{i\chi_n \hat{\cal A}} e^{i\varphi_k \hat J_z} U_\sigma$ in the $N_s$-dimensional single-particle space. In particular, $e^{i\varphi_k \hat j_z}$ is a diagonal matrix with elements $e^{i\varphi_k m_a}$  ($m_a$ is the magnetic quantum number of orbital $a$). The canonical $M$-projected partition ${\rm Tr}_M U_\sigma$ is calculated from Eqs.~(\ref{M-project}) and (\ref{canonical-trace}).

Similarly, the canonical expectation value of an observable $O$ at fixed $M$ is  calculated from
\begin{equation}\label{M-observ}
\langle O \rangle_M  \equiv {{\rm Tr}_M \left( O e^{-\beta H} \right) \over {\rm Tr}_M e^{-\beta H} }
={\left\langle {{\rm Tr}_M
(O U_\sigma)  \over {\rm Tr} U_\sigma} \Phi_\sigma \right\rangle_W
\over \left\langle { {\rm Tr}_M U_\sigma  \over
{\rm Tr}\, U_\sigma} \Phi_\sigma \right\rangle_W
} \;,
\end{equation}
where ${\rm Tr}_M \,U_\sigma$ is given by (\ref{M-project}), and
\begin{equation}\label{M-observable}
 \!\!\! \! {\rm Tr}_M\,
(O U_\sigma)  = {1\over 2J_s +1} \sum\limits_{k=-J_s}^{J_s} e^{-i \varphi_k M}
     {\rm Tr}\, (O e^{i\varphi_k \hat J_z} U_\sigma) \;.
\end{equation}
The canonical trace on the r.h.s.\ of (\ref{M-observable})  can be calculated by particle-number projection.  For example, for a one-body operator $O= \sum_{ab} \langle a| O |b\rangle a^\dagger_a a_b $ we find an expression similar to Eq.~(\ref{canonical-trace}), but each term in the sum includes the additional factor
\begin{equation}
{{\rm Tr}\, (a^\dagger_a a_b e^{i\chi_n \hat{\cal A}}e^{i\varphi_k \hat J_z} U_\sigma)
\over  {\rm Tr}\, (e^{i\chi_n \hat{\cal A}} e^{i\varphi_k \hat J_z} U_\sigma)}
= \left({ {\bf I}  \over {\bf I} + {{\bf U}^{(n,k)}_\sigma}^{-1}}\right)_{ba} \;,
\end{equation}
where here the traces are grand-canonical.

 The spin-projected partition function at fixed total spin $J$ is defined by $Z_J(\beta) \equiv {\rm Tr}_J e^{-\beta H} = \sum_{\alpha} \langle \alpha J M | e^{-\beta H} | \alpha J M \rangle =\sum_\alpha e^{-\beta E_{\alpha J}}$
and is independent of $M$.
The $J$-projected partition can be expressed as a difference of corresponding $M$-projected partitions
\begin{equation}\label{J-projection}
{\rm Tr}_J  e^{-\beta H} = {\rm Tr}_{M=J}  e^{-\beta H} - {\rm Tr}_{M=J+1}  e^{-\beta H}  \;.
\end{equation}
Eq. (\ref{J-projection}) holds since $ e^{-\beta H}$ is a {\it scalar} operator. Using the HS representation (\ref{M-ratio}) for both $M=J$ and $M=J+1$, we find
\begin{equation}\label{J-ratio}
{Z_J(\beta) \over Z(\beta)} = { \left\langle \left({{\rm Tr}_{M=J} U_\sigma \over {\rm  Tr} U_\sigma}- {{\rm Tr}_{M=J+1} U_\sigma \over {\rm  Tr} U_\sigma}\right) \Phi_\sigma \right\rangle_W  \over \langle \Phi_\sigma \rangle_W}
\;,
\end{equation}
where ${\rm Tr}_M U_\sigma$ are calculated as before. It is also possible to apply the HS transformation directly in $Z_J$ and obtain
${Z_J(\beta)/ Z(\beta)} = {\left\langle \left({{\rm Tr}_J U_\sigma / {\rm  Tr} U_\sigma}\right) \Phi_\sigma \right\rangle_W / \langle \Phi_\sigma \rangle_W}$. This relation is not equivalent to  Eq.~(\ref{J-ratio}) since $U_\sigma$ is not rotationally invariant and ${\rm Tr}_J U_\sigma \neq {\rm Tr}_{M=J} U_\sigma  - {\rm Tr}_{M=J+1}U_\sigma$. However, the calculation of ${\rm Tr}_J U_\sigma $ requires a full spin projection and is considerably more time-consuming than the $M$ projection required in (\ref{J-ratio}).

To calculate the spin-projected expectation value $\langle O \rangle_J  \equiv {{\rm Tr}_J \left( O e^{-\beta H} \right) / {\rm Tr}_J e^{-\beta H} }$ of  a {\em scalar} observable $O$ (e.g., the energy), we note that
${\rm Tr}_J(O e^{-\beta H}) = {\rm Tr}_{M=J} (O e^{-\beta H})  - {\rm Tr}_{M=J+1} (O e^{-\beta H})$. Applying the HS transformation, we find
\begin{equation}\label{J-observ}
\langle O \rangle_J
 = { \left\langle \left({{\rm Tr}_{M=J}(O U_\sigma) \over {\rm  Tr} U_\sigma}- {{\rm Tr}_{M=J+1}(O U_\sigma) \over {\rm  Tr} U_\sigma}\right) \Phi_\sigma \right\rangle_W  \over
\left\langle \left({{\rm Tr}_{M=J} U_\sigma \over {\rm  Tr} U_\sigma}- {{\rm Tr}_{M=J+1} U_\sigma \over {\rm  Tr} U_\sigma}\right) \Phi_\sigma \right\rangle_W }
\;,
\end{equation}
where $M$-projected quantities are calculated as before.

For a good-sign interaction, $U_\sigma$ is time-reversal invariant.
Since  $e^{i\varphi_k J_z}$ is always time-reversal invariant,
so is $e^{i\varphi_k J_z} U_\sigma$, 
and its grand-canonical trace is always positive
(since the eigenvalues of the single-particle matrix
$e^{i\varphi_k j_z} {\bf U}_\sigma$ come in complex conjugate pairs).
When projected on an even number of particles,
${\rm Tr}\,(e^{i\varphi_k J_z} U_\sigma)$ remains almost always
positive.
In Eq.~(\ref{M-project}) we are summing positive numbers
with coefficients $e^{-i\varphi_k M}$,
leading to $M$-projected partition ${\rm Tr}_M U_\sigma$
that can be non-positive,
with the exception of the $M=0$ case.
However, this sign problem becomes severe only above a certain $\beta$,
and typically occurs at smaller values of $\beta$ as $M$ gets larger.
We encounter a similar situation for the $J$ projection
with the $J=0$ projection having no sign problem.
In practice, the level density at higher spin values
becomes appreciable only at higher excitations,
and meaningful spin distributions can be extracted except
for very low excitations.

\begin{figure}[h!]
\epsfxsize= 0.9\columnwidth \centerline{\epsffile{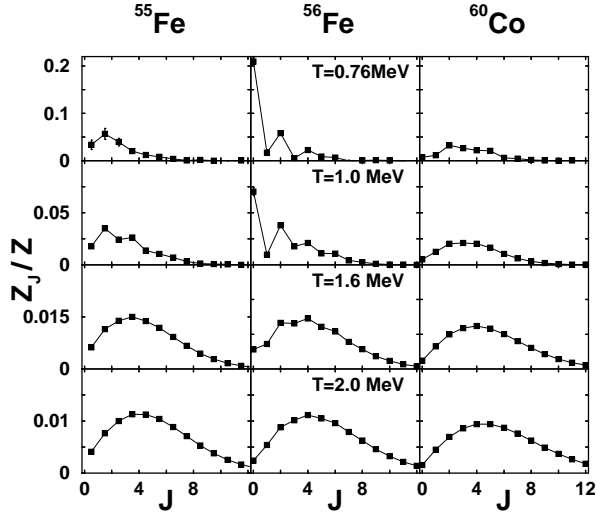}}
\caption {SMMC spin distributions of the partition
function $Z_J/Z$ for $^{55}$Fe (left panels), $^{56}$Fe (middle), and $^{60}$Co (right).
}
\label{zj-ratio}
\end{figure}

We used the spin projection method to calculate the spin distribution of the partition function and level density in the presence of correlations. In particular we calculated such spin distributions for $^{56}$Fe (an even-even nucleus), $^{55}$Fe (odd-even), and $^{60}$Co (odd-odd) in the complete $(pf+0g_{9/2})$-shell, and for $\beta$ in
the range from 0 to $\sim 2$~MeV$^{-1}$ using the Hamiltonian of Ref.~\cite{na97}. The SMMC
results for $Z_J/Z$, calculated from Eq.~(\ref{J-ratio}), are shown in
Fig.~\ref{zj-ratio}. For temperatures $T =\beta^{-1} \lesssim 1.5$~MeV,
an odd-even staggering is observed in the even-even nucleus $^{56}$Fe.
In particular, $Z_{J=0}/Z$ is significantly enhanced as $T$ decreases.
No odd-even spin staggering effect is observed in the odd-even and odd-odd nuclei.

\begin{figure}
\epsfxsize= 0.9 \columnwidth \centerline{\epsffile{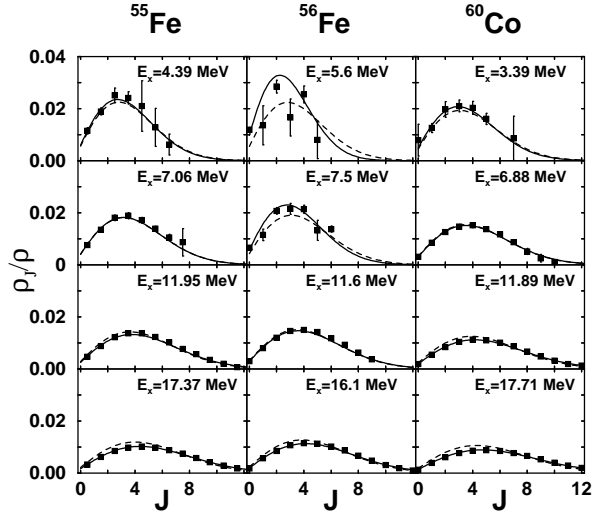}}
\caption { Spin distribution of level densities, $\rho_J/\rho$,
at constant excitation energy $E_x$ for $^{55}$Fe, $^{56}$Fe and $^{60}$Co.
The SMMC results (solid squares) are compared with the spin-cutoff model (\protect\ref{spin-cutoff}) with $\sigma^2$ fitted to the SMMC results
(solid lines), and with $\sigma^2$ calculated from Eq.~(\protect\ref{sigma-I}) using the rigid-body moment of inertia (dashed lines).
 }
 \label{spin-dist}
\end{figure}

We calculated spin-projected thermal energies $\langle H\rangle_M$ and
$\langle H \rangle_J$ as a function of $\beta$ and used the method
of Refs.~\cite{na97,al99}
to obtain the level densities $\rho_M(E_x)$ and $\rho_J(E_x)$ as a function of excitation energy $E_x$. The total level density $\rho(E_x)$ was found from $\langle H\rangle$, so we could determine the spin distribution $\rho_J/\rho$  at fixed values of the excitation energy.  In Fig.~\ref{spin-dist} we show the spin distribution of $\rho_J / \rho$
at several excitation energies for  $^{56}$Fe (middle panels), $^{55}$Fe (left panels), and $^{60}$Co (right panels). The solid squares are the SMMC results, while the solid lines describe fits (at fixed $E_x$) to the spin-cutoff model
\begin{equation}
  \label{spin-cutoff}
  \rho_J(E_x) = \rho(E_x)
  {(2J+1) \over 2\sqrt{2 \pi} \sigma^3} e^{-{J(J+1) \over 2
      \sigma^2}}\;,
\end{equation}
with an energy-dependent spin-cutoff parameter $\sigma$ as the only fit parameter. The spin-projected density $\rho_J(E_x)$ in (\ref{spin-cutoff}) is normalized such that $\sum_J (2J+1) \rho_J(E_x)
\approx \rho(E_x)$.
Equation~(\ref{spin-cutoff}) follows from the random coupling model,
in which the distribution of the total spin vector ${\bf \hat J}$ is Gaussian~\cite{er60}.  At intermediate and high excitation energies the spin-cutoff model seems to work well for all three
nuclei.  However, for the even-even
nucleus $^{56}$Fe, we observe an odd-even (spin) staggering effect below
$E_x\sim 8$ MeV that cannot be explained by the spin-cutoff model.

The energy-dependent spin-cutoff parameter $\sigma^2(E_x)$, obtained by fitting  $\rho_J/\rho$ to Eq.~({\ref{spin-cutoff}), is shown (solid squares) versus $E_x$ in the top panels of Fig.~\ref{sig-inertia} for $^{55}$Fe, $^{56}$Fe, and $^{60}$Co.
The quantity $\sigma^2$ can also be obtained from fits to ${\rho_M
/\rho}$ [in the spin-cutoff model  ${\rho_M / \rho}= (2 \pi
\sigma^2)^{-1/2}{e^{-M^2/2 \sigma^2}}$] but the results are similar.
Despite the deviation from (\ref{spin-cutoff})
at $E_x\lesssim 8$\,MeV in $^{56}$Fe,
the fitted $\sigma^2(E_x)$ represents well the average behavior of
$\rho_J/\rho$.
\begin{figure}
\epsfxsize= 0.9 \columnwidth \centerline{\epsffile{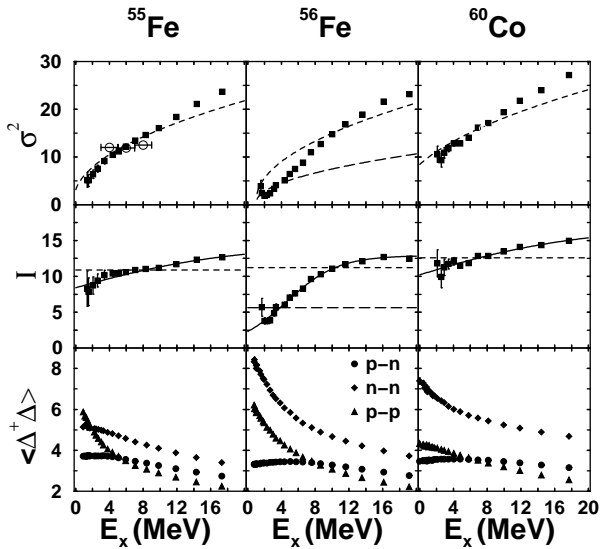}}
\caption {
Shown from top to bottom are the spin-cutoff parameter $\sigma^2$
 (extracted from the spin distributions),
the moment of inertia $I$ in (\protect\ref{sigma-I}) and
the number of $J=0$ pairs $\langle \Delta^\dagger \Delta \rangle$
 for $^{55}$Fe (left panels), $^{56}$Fe (middle), and $^{60}$Co
 (right).
The SMMC results for $\sigma^2$ and $I$ are denoted by solid squares.
The open circles with horizontal errors in the $\sigma^2$ panel of
 $^{55}$Fe
are experimental data~\protect\cite{gr74}.
The dashed lines correspond to a rigid-body moment of inertia
while the long dashed lines (for $^{56}$Fe only) correspond to
 half the rigid-body moment of inertia.
  }
  \label{sig-inertia}
\end{figure}

There is not much data available regarding the spin-cutoff parameter.  A few available experimental data points are shown for $^{55}$Fe~\cite{gr74}. The SMMC calculations are in general agreement with the experimental data.

$\sigma^2$ is related to an effective moment of inertia $I$ through
\begin{equation}
  \label{sigma-I}
  \sigma^2 = { I T \over \hbar^2} \;,
\end{equation}
where $T$ is the temperature.
Using Eq.~(\ref{sigma-I}) we can convert the SMMC values of $\sigma^2$ to
an energy-dependent moment of inertia $I(E_x)$.
The results are shown in the middle panels of Fig.~\ref{sig-inertia} (solid squares). For comparison we also show the
rigid-body value $ I/\hbar^2 = 0.0137 A^{5/3}$ MeV$^{-1}$
(dashed lines), and  (for $^{56}$Fe only) half the rigid body value (long dashed lines) of the moment of inertia. In all three nuclei,
$I(E_x)$ is a monotonically increasing function of $E_x$ and is close to the rigid-body value at intermediate and high excitations. However, for energies below $\sim 8 - 10$ MeV we observe a suppression that is particularly strong for the even-even nucleus $^{56}$Fe.

The energy dependence of the moment of inertia extracted from the spin distributions originates in pairing correlations. To demonstrate that we calculated the average number of $J=0$  nucleon pairs $\langle \Delta^\dagger \Delta \rangle$, where
$\Delta^\dagger =
\sum_{a m_a>0}
(-1)^{j_a - m_a}
a^{\dagger}_{j_a m_a}a^{\dagger}_{j_a -m_a}$.
 The SMMC results for proton (p-p), neutron (n-n) and proton-neutron (p-n) pairs are shown versus $E_x$ in the bottom panels of Fig.~\ref{sig-inertia}. The rapid decrease of the number of $p$-$p$ and $n$-$n$ pairs for $^{56}$Fe is strongly correlated with the rapid
increase observed of the moment of inertia.  The correlation between $I$ and the number of
pairs suggests that at low excitation energy the nucleons behave
as condensed BCS pairs, leading to moment of inertia values that
are significantly smaller than the rigid-body value. Thus the spin distributions provide a thermal signature of the pairing phase transition. Thermal signatures of pairing correlations were previously observed in the heat capacity~\cite{li01, al03}. The strong odd-even effect seen in the moment of inertia can be explained by a simple pairing model plus a number parity projection~\cite{al05}.
\begin{figure}[h!]
\epsfxsize= 0.65\columnwidth \centerline{\epsffile{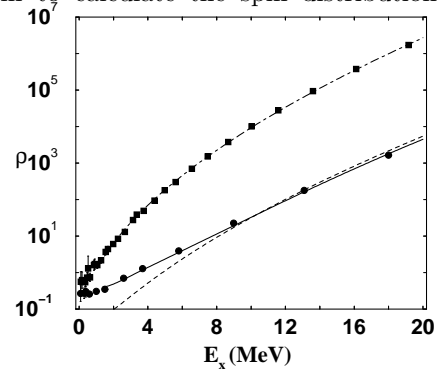}}
\caption{Total and  $J=0$ level densities for $^{56}$Fe. The total SMMC level density (solid squares) is well described by the BBF level density (dotted-dashed line). The $J=0$ SMMC level density (solid circles) is compared to the $J=0$ level density inferred from (\protect\ref{spin-cutoff}) with the fitted moment of
inertia (solid line) and the rigid-body moment of inertia (dashed line). The suppression of the moment of inertia at low excitations enhances the $J=0$ level density $E_x \alt 8$ MeV.}
\label{j0-density}
\end{figure}

The strong suppression of the moment of inertia at low
excitations for even-even nuclei has a signature in the $J=0$ level
density. In Fig.~\ref{j0-density} we show the total and
$J=0$ level densities for $^{56}$Fe. The SMMC total density (solid squares) is in good agreement with the backshifted Bethe formula (BBF)
(dotted-dashed line). The SMMC $J=0$ level
density is shown by the solid circles. The solid line describes the $J=0$ level density obtained from (\ref{spin-cutoff}) (and the BBF for the total $\rho$) using the energy-dependent
moment of inertia in Fig.~\ref{sig-inertia}. The dashed line
is also from Eq.~(\ref{spin-cutoff}) but with a rigid-body moment of
inertia. The rigid-body curve agrees with the SMMC results at high
excitations, but shows deviations for energies below
$\sim 8$~MeV. The enhancement of the $J=0$ level density for $E_x \lesssim 8$~MeV reflects the decrease of the moment of inertia due to pairing correlations.

In conclusion, we have introduced spin projection methods in the shell model Monte Carlo approach and used them to calculate the spin distributions of level densities. The energy-dependent moment of inertia extracted from these distributions displays an odd-even effect that is a signature of the pairing correlations.

This work was supported in part by the U.S. DOE grant No.\
DE-FG-0291-ER-40608 and the Grant-in-Aid for Scientific Research (B), No. 15340700, by the
MEXT, Japan. Computational cycles
were provided in part by
the NERSC high performance computing facility at LBL.
We thank G.F. Bertsch for helpful discussions.

\end{document}